\documentclass[fleqn,usenatbib]{mnras}

\usepackage{newtxtext,newtxmath}
\usepackage[T1]{fontenc}

\DeclareRobustCommand{\VAN}[3]{#2}
\let\VANthebibliography\thebibliography
\def\thebibliography{\DeclareRobustCommand{\VAN}[3]{##3}\VANthebibliography}

\usepackage{graphicx}
\usepackage{amsmath}
\usepackage{xspace}
\usepackage{lineno}
\usepackage[dvipsnames]{xcolor}

\input{hyperlink-year-only-natbib-patch}

\newcommand{\SatGen}{\textsc{SatGen}\xspace}

\newcommand{\unit}[1]{\ensuremath{\mathrm{\,#1}}\xspace}
\newcommand{\Msun}{\unit{M_\odot}}

\title[Bound fractions across cosmo sims]{Simulations agree: the Milky Way's satellites are disrupting}

\author[A.~H.~Riley et al.]{Alexander~H.~Riley,$^{1,2}$\thanks{E-mail: alexander.riley@fysik.lu.se (AHR)}
Nora Shipp,$^{3}$
J.~Sebastian Monzon,$^{4}$
Arpit Arora,$^{3}$
Adriana Dropulic,$^{5}$
\newauthor Gray Thoron,$^{3}$
Oscar Agertz,$^{1}$
and Santi Roca-F\`{a}brega$^{1,6}$
\\
$^{1}$Lund Observatory, Division of Astrophysics, Department of Physics, Lund University, SE-221 00 Lund, Sweden \\
$^{2}$Institute for Computational Cosmology, Department of Physics, Durham University, South Road, Durham DH1 3LE, UK \\
$^{3}$Department of Astronomy, University of Washington, Seattle, WA 98195, USA \\
$^{4}$Department of Astronomy, Yale University, New Haven, CT 06520, USA \\
$^{5}$DARK, Niels Bohr Institute, University of Copenhagen, Jagtvej 155, 2200 Copenhagen, Denmark \\
$^{6}$Departamento de F\'{i}sica de la Tierra y Astrof\'{i}sica, Facultad de Ciencias F\'{i}sicas, Plaza Ciencias, 1, 28040 Madrid, Spain
}

\date{Accepted XXX. Received YYY; in original form ZZZ}

\pubyear{\the\year{}}

\begin{document}
\label{firstpage}
\pagerange{\pageref{firstpage}--\pageref{lastpage}}
\maketitle

\begin{abstract}
Recent analyses of simulated Milky Way-mass galaxies have highlighted the prevalence of satellites undergoing tidal disruption.
However, these have focused on single hydrodynamical codes with a number of limitations, making it unclear how generic this result is.
We compare satellite disruption in four models: two cosmological hydrodynamical simulations (Auriga and FIRE), cosmological $N$-body simulations with stellar particle tagging (Aquarius), and a semi-analytic model with varying host disk properties (\SatGen).
For all systems, we compute the fraction of stellar mass retained by the progenitor ($f_\text{bound}$) at the present day.
We find a remarkable agreement between FIRE, Aquarius, and \SatGen; all predict that $\sim$20 to 30~per~cent of classical-mass satellites have $f_\text{bound} < 0.9$ ($\sim$5 to 10~per~cent have $f_\text{bound} < 0.5$).
Satellites in Auriga experience substantially more mass loss than other simulations, which we attribute to their massive central galaxies and large satellite sizes prior to infall.
In the ultra-faint regime captured by Aquarius and \SatGen, disruption rates are comparable to or exceeding those for classical satellites.
The ubiquity of mass loss across models presents both an exciting opportunity (a plethora of stellar streams awaiting discovery) and a nuisance (for theorists using satellite counts to constrain dark matter).
\end{abstract}


\section{Introduction}

Satellite galaxies surrounding Milky Way-mass hosts have been extensively studied in cosmological simulations \citep[e.g.,][]{Libeskind:2007, Okamoto:2010, Brooks:2014,Sawala:2016, Wetzel:2016, Applebaum:2021} and semi-analytic models \citep{Kauffmann:1993, Bullock:2000, Benson:2002, Kim:2018, Kim:2026, Nadler:2020mwsats, Manwadkar:2022, Ahvazi:2025} to resolve small-scale challenges to the prevailing $\Lambda$CDM framework \citep{Sales:2022} or use their observed properties to constrain the nature of dark matter \citep{Nadler:2026}.
Recent analyses of cosmological hydrodynamical simulations have shown that these satellites are often actively losing stellar mass, forming extended stellar streams \citep{Wang:2017, Panithanpaisal:2021, Riley:2025, Pathak:2025, Meziani:2026}.
While this may be surprising, given the paucity of tidal tails observed around Milky Way satellites, this debris may be hidden beneath the surface brightness limits of current wide-area photometric surveys \citep{Wang:2017, Shipp:2023}.\footnote{Detections of clear tidal tails from Antlia~II and Crater~II \citep{Ji:2021, Atzberger:2026, Limberg:2026} have helped close this gap, but a broad mismatch between simulations and observations remains.}

It is unclear to what degree these disruption rates are simply reflecting theoretical limitations, either \textit{numerical} shortcomings of the simulations or improper \textit{physical} properties of the satellites.
Tailored $N$-body simulations have highlighted that subhaloes can disrupt artificially due to numerical noise from insufficient number of particles or force resolution (\citealp{vandenBosch:2018, Errani:2020, Errani:2021, Chiang:2026}; Reddy et al.~in prep).
Spurious heating due to unequal dark matter and star particle masses can result in galaxies expanding in size (\citealp{Ludlow:2019, Ludlow:2023}; Liao et al.~submitted), possibly enhancing disruption upon infall to a larger host.
Even beyond these numerical effects, simulated systems that do not follow established scaling relations (e.g., stellar mass-halo mass, stellar mass-size) could experience physical disruption that is nonetheless unrealistic (\citealp{Penarrubia:2008}; Reddy et al.~in prep).
On the other hand, many of the careful numerical experiments previously mentioned do not capture physical mechanisms that enhance tidal disruption present in cosmological setups, including preprocessing by massive satellites \citep{He:2025}, the tidal field of a growing host with a central disk \citep{Garrison-Kimmel:2017, Kelley:2019, Grand:2021, Wang:2025}, and satellite-satellite collisions \citep{Shipp:2025}.

These theoretical uncertainties have led to repeated calls to study satellite disruption across a variety of models \citep[see discussion in][]{Shipp:2023, Shipp:2025}.
One principled approach to this task could be to analyze disrupting satellites in haloes run with the same initial conditions using different numerical prescriptions, in the spirit of the AGORA comparison project \citep{Kim:2014, Jung:2024}.
However, this may limit the types of models that can be considered, excluding semi-analytic approaches that are not limited by particle resolution.
Another approach is to take a variety of available simulations, with their respective strengths and weaknesses, and invest effort to standardize the analysis as much as possible.
If a similar result emerged across methods, this would suggest that the high rate of stellar mass loss is not an effect of one theoretical framework or numerical artefact.

In this work we take the latter approach, comparing disrupting satellites in four different publicly available simulation suites: Auriga, FIRE, Aquarius, and \SatGen.
These models span a variety of galaxy-halo connections for populating satellites, numerical resolution limitations, and models for enhanced tidal disruption from the host galaxy.
In Section~\ref{sec:data} we introduce key aspects of the simulation suites and the corresponding catalogues of accreted satellites.
In Section~\ref{sec:methods} we describe how we compute $f_\text{bound}$ for the simulations and how we homogenise the merger catalogues.
We present our main results in Section~\ref{sec:results} and discuss their implications in Section~\ref{sec:conclusions}.

\section{Simulations} \label{sec:data}

\subsection{Auriga}

The Auriga simulations \citep{Grand:2017, Grand:2024} are cosmological magnetohydrodynamical zoom simulations of isolated Milky Way-mass galaxies.
The haloes were selected from the EAGLE dark-matter-only box \citep{Schaye:2015} to have $M_\text{200c} = 1-2\times10^{12}$M$_\odot$ and satisfy an isolation criterion.
They were resimulated using \textsc{Arepo} \citep{Springel:2010, Pakmor:2016} with a galaxy formation model detailed in \citet{Grand:2017, Grand:2024}.
We use the six Auriga haloes that were simulated at `level~3' resolution, with baryonic element (dark matter particle) masses of $6.7(36)\times10^3$~M$_\odot$ and a minimum softening length of 188~pc.
Halo catalogues were constructed using the SUBFIND halo finder \citep{Springel:2001} and connected across time using the LHaloTree merger tree algorithm \citep{Springel:2005}.

To identify accreted systems and isolate their constituent star particles, we use the `accreted particle lists' in the Auriga public data release \citep{Grand:2024}.
To construct these catalogues, star particles are assigned to the (sub)halo they are bound to immediately following their birth.
The union of star particles that formed in the main branch of a halo across cosmic time correspond to one system (practically, such particles are assigned the same \texttt{PeakMassIndex}).
These are the same catalogues that are used in \citet{Riley:2025} and \citet{Shipp:2025}.

\subsection{FIRE}

We use isolated Milky Way-mass haloes from the Latte suite \citep{Wetzel:2016}, run as part of the Feedback in Realistic Environments (FIRE) project.
The simulations were run with \textsc{Gizmo} \citep{Hopkins:2015} using the FIRE-2 galaxy formation model \citep{Hopkins:2018}.
We use the set of `m12' haloes listed in Table~1 of \citet{Wetzel:2023}, originally presented across a number of studies \citep{Wetzel:2016, Garrison-Kimmel:2017, Garrison-Kimmel:2019, Hopkins:2018, Samuel:2020}.
We use seven haloes in total, excluding m12z as its merger catalogue is not available.
The haloes were simulated with baryonic (dark matter) mass resolution of $7.1(35)\times10^{3}$~M$_\odot$\footnote{The m12r and m12w haloes were run at slightly worse dark matter resolution of 39,000\Msun \citep{Samuel:2020}.} and a softening length of 1 (40)~pc.
Halo catalogues are produced using the \textsc{Rockstar} algorithm \citep{Behroozi:2013rockstar} and linked into a merger tree using \textsc{consistent-trees} \citep{Behroozi:2013consistenttrees}.

To identify stars in each accretion event, we use catalogues assembled in the same manner as \citet{Panithanpaisal:2021}.
They first construct a list of potential stellar stream candidates by identifying satellites between 2.7 and 6.5~Gyr ago whose star particles are within the host's $r_\text{200m}$ at the present day.
To account for satellites that accreted more recently, they augment these catalogues with all present-day satellites (within the host's $r_\text{200m}$) that were not accounted for in the previous step.
Using the merger trees, they identify the snapshot that each object obtained its maximum stellar mass.
They then collect all star particles assigned\footnote{When assigning star particles to a halo, \citet{Panithanpaisal:2021} select all star particles located within the halo's virial radius with relative velocities less than twice that halo's circular velocity curve.} to this object for snapshots within a 200~Myr window of this snapshot.

\subsection{Aquarius}

\begin{figure}
    \centering
    \includegraphics[width=\linewidth]{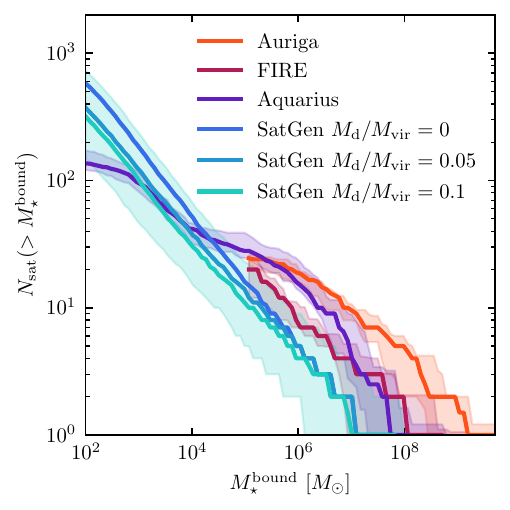}
    \caption{The stellar mass functions of satellite galaxies across all simulations considered in this work.
    Shaded contours correspond to the 1$\sigma$ scatter across haloes, while the solid lines are the medians.
    For visual clarity, we only show the \SatGen scatter for the $M_\text{d} / M_\text{vir} = 0.1$ model, which is similar across disc scenarios.
    \textit{The models span a variety of galaxy-halo connections and host properties, resulting in different stellar mass functions for satellites.}
    }
    \label{fig:luminosity-functions}
\end{figure}

The Aquarius simulations \citep{Springel:2008} are cosmological zoom simulations of isolated Milky Way-mass haloes.
The six haloes were selected from a lower-resolution version of the Millennium-II simulation \citep{Boylan-Kolchin:2009} and resimulated with \textsc{Gadget-3}, an updated version of \textsc{Gadget-2} \citep{Springel:2005gadget}.
To produce stellar distributions for these dark-matter-only simulations, \citet{Cooper:2010} used the GALFORM semi-analytic model \citep{Cole:2000, Bower:2006} to model star formation histories for each galaxy and the STINGS particle tagging framework \citep{Cooper:2017, Cooper:2025} to assign stars to dark matter particles (for brevity, we refer to these outputs as `Aquarius').
The resulting star particle data were released publicly by \citet{Pu:2025}.

\citet{Cooper:2010} tagged Aquarius haloes that were simulated at `level 2' resolution, with dark matter particle masses ranging from $6.4-14\times10^3$~M$_\odot$ and a softening length of 65.8~pc.
Halo catalogues were constructed using the SUBFIND halo finder \citep{Springel:2001} and connected across time using the \textsc{D-Trees} merger tree algorithm \citep{Jiang:2014dhaloes}.
Stars formed in a given halo (according to GALFORM) are assigned to that halo's most tightly bound particles\footnote{\citet{Cooper:2010} select the most-bound one per cent of particles each snapshot, with stellar mass shared equally among those selected.} in the following snapshot.
By tracking tagged particles to the present day, accreted stellar haloes comprised of satellites, stellar streams, and phase-mixed debris naturally emerge.

To identify accretion events, we use the progenitor catalogues described in \citet{Pu:2025}.
Star particles that become unbound from a given satellite to join the Milky Way-mass host are assigned the same \texttt{ProgenitorID}.
Stars that remain bound to the progenitor at the present day are also assigned that corresponding ID.

\subsection{SatGen}

The \SatGen code \citep{Jiang:2021, Green:2021} is a semi-analytic framework to generate statistical samples of satellite galaxy populations.
Briefly, \SatGen constructs merger trees for dark matter host halos in a given cosmology using the extended Press-Schechter (EPS) method of \citet{Parkinson:2008}.
Once the trees are constructed, \SatGen assigns stellar masses \citep[see Appendix J of][]{Behroozi:2019} and sizes to progenitor halos at accretion and integrates their orbits accounting for dynamical friction and tidal stripping using prescriptions that are calibrated against a large suite of high-resolution idealized simulations.

Throughout the process, \SatGen tracks how much stellar mass each satellite loses using tidal tracks defined in \cite{Errani:2018} which assume that the ratios stellar mass ($m_{*}(t) / m_{*}$) and stellar size ($r_{\rm eff}(t) / r_{\rm eff}$) can be characterized by
\begin{equation}
    g(x) = \left(\frac{x_s + 1}{x_s + f_b}\right)^{\mu} f_b^{\eta}\,.
    \label{eq:errani}
\end{equation}
Here, $g(x)$ refers to either $m_{*}(t) / m_{*}$ or $r_{\rm eff}(t) / r_{\rm eff}$, and $f_b$ is the instantaneous bound mass fraction of the corresponding dark matter subhalo.
The additional parameters $\mu$, $\eta$ and $x_s$ depend on the `cuspiness' of the subhalo at accretion, parameterized by the inner logarithmic slope of the subhalo's density profile, and on the ratio $r_{\rm eff}/r_{\rm max}$.
Here, $r_{\rm max}$ is the radius at which the subhalo reaches its maximum circular velocity, $v_{\rm max}$, which for an unevolved NFW density profile is $\sim$2.16~$r_{\rm s}$.
\SatGen uses a simple interpolation scheme to determine the values of $\mu$, $\eta$ and $x_s$ based on the values inferred by \cite{Errani:2018} using a discrete set of simulation results that explored the relevant parameter space.\footnote{Since those simulations modeled the stellar bodies as Plummer spheres, we are implicitly assuming the same.}

We use merger trees processed in the same manner as in \citet{Monzon:2024, Monzon:2026} that fall in a host mass range $M_\text{vir} = 10^{11.5} - 10^{12.5}$.
At every time step we assume that the host's dark matter potential follows an NFW profile \citep{Navarro:1997}.
Halo concentrations are computed using the model from \cite{Zhao:2009}.
The galactic disc, when included, is positioned at the centre of the host halo and modeled with the axisymmetric \citet{Miyamoto:1975} disk that grows with the host (as a constant fraction of the host's $M_\text{vir}$, see \citealp{Green:2021}).
We consider three scenarios: no embedded disc potential, a disc with $M_\text{d} / M_\text{vir} = 0.05$, and a disc with $M_\text{d} / M_\text{vir} = 0.1$.
Each sample contains 500 unique accretion histories (specifically, a subset of those used in \citealp{Dropulic:2025}) that are identical across the three disc scenarios.

\section{Methods} \label{sec:methods}

\subsection{Homogenising the merger catalogues}

While each of our suites produce catalogues that have comparable contents (a list of accreted systems and their associated star particles), they were not constructed in exactly the same way.
To facilitate our $f_\text{bound}$ comparison, we apply the following selection criteria to all catalogues:
\begin{enumerate}
    \item Satellite: has a bound component at the present day
    \item Mass: total stellar mass $M_\star > 5\times10^5 \Msun$ and bound stellar mass $M_\star^\text{bound} > 10^5 \Msun$
    \item Infall time: fell into the main host within the past 10~Gyr
\end{enumerate}

The first two criteria ensure we are studying surviving classical-mass satellites that are resolved in both Auriga and FIRE.
The final one alleviates differences in how far back in time objects need to be tracked.
We also consider ultra-faint satellites in Aquarius and \SatGen, which capture the tidal evolution of much fainter systems than the hydrodynamical suites.
For this, we consider systems with $10^2 \Msun < M_\star^\text{bound} < 10^5 \Msun$ instead of the mass criteria listed above.
We do not apply a selection in distance from the main host.

\subsection{Bound fractions} \label{sec:bound-fraction-calc}

\begin{figure*}
    \centering
    \includegraphics[width=\linewidth]{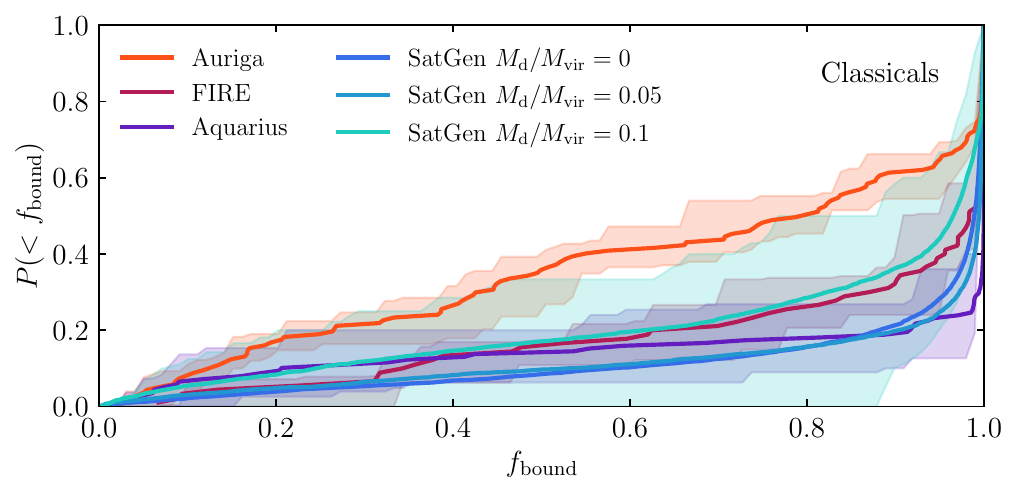}
    \caption{The cumulative distribution of $f_\text{bound}$ for classical-mass satellites across different simulation setups.
    Simulations with satellite systems that are the least disrupted overall will be towards the lower-right corner.
    We only include systems that have a total stellar mass $M_\star > 5\times10^5$~M$_\odot$ and a bound stellar mass of $M_\star^\text{bound} > 10^5$~M$_\odot$.
    Shaded contours correspond to the 1$\sigma$ scatter across haloes, while the solid lines are the result considering all systems `stacked'.
    \textit{All of the models predict substantial stellar mass loss, with three (FIRE, Aquarius, and \SatGen) further agreeing on the amount (within host-to-host scatter).}
    }
    \label{fig:fbound-classicals}
\end{figure*}

Standard halo finders can miss disrupting subhaloes at the resolution typical of zoom simulations \citep{Han:2018, Diemer:2024, Mansfield:2024}.
To alleviate these concerns, we apply a novel iterative unbinding approach (implemented in \texttt{nbody\_streams}; \citealp{Arora:2026}) to both the Auriga and FIRE catalogues.
For each snapshot after falling into the main host, we first determine the center of the satellite using a mass-weighted kernel density estimator (KDE) applied to the star particles.
The mean position of particles within the highest 1~per~cent of the KDE is taken as the initial center.
We then compute the potential of the satellite's dark matter and star particles using a multipole Poisson solver implemented in \textsc{Agama} \citep{Vasiliev:2019}.
Particles not bound to this potential are temporarily masked and the potential re-computed, iterating until the number of bound particles changes by less than 0.1~per~cent.
Satellites are tracked forwards in time until fewer than 5 bound particles remain.

This technique results in modestly larger bound fractions at intermediate stages of disruption ($0.2 < f_\text{bound} < 0.8$) compared to simply relying on the halo finder (see the Appendices of \citealp{Thoron:2026} for Auriga and \citealp{MandacaruGuerra:2026} for FIRE).
We note that this method does not compensate for numerical disruption due to low particle count or unequal particle masses \citep{vandenBosch:2018, Ludlow:2019, Ludlow:2023}.
If a subhalo disrupts earlier than it should, there is nothing for an improved halo tracker to find.

For the other simulations, $f_\text{bound}$ is computed in a more straightforward manner.
We assume that the high resolution of Aquarius mitigates concerns of SUBFIND's ability to track the evolution of subhaloes massive enough to form galaxies.\footnote{\citet{Santos-Santos:2025, Santos-Santos:2026} raise concerns about the ability of SUBFIND and GALFORM to track the faintest satellites in Aquarius, even at higher resolution than considered here. We remind that satellites with $M_\star^\text{bound} > 10^5 \Msun$ are the primary focus of this work.}
Tidal mass loss in \SatGen is based on calibrated models (for dark matter) and tidal tracks (for stars), such that identifying the bound remnant is not a concern.

\section{Results} \label{sec:results}

\subsection{Stellar mass functions}

In Figure \ref{fig:luminosity-functions} we show the stellar mass functions for satellite galaxies across all models considered in this work.
These include the iterative unbinding corrections to Auriga and FIRE described in Section~\ref{sec:bound-fraction-calc}.
We intentionally restrict the results from the hydrodynamical simulations (Auriga and FIRE) to $M_\star^\text{bound} > 10^5 \Msun$, reflecting their resolution limitations.
We caution that the number of satellites is sensitive to host halo mass and assembly history \citep[e.g.,][]{Boylan-Kolchin:2010, Joshi:2024}, and it is not our goal to evaluate how well these predictions compare to observed luminosity functions.

Overall the predictions span a broad range at a given stellar mass, typically over 0.5~dex.
Auriga, FIRE, and Aquarius are in reasonable agreement on the typical number of classical satellites ($M_\star^\text{bound} > 10^5 \Msun$), all falling between 20 to 30.
Auriga has the largest number of massive ($M_\star^\text{bound} > 10^7 \Msun$) satellites, with far fewer classical satellites below this mass.
This reflects the high stellar mass-halo mass relation in Auriga relative to other simulations, particularly for haloes with $M_\text{peak} \sim 10^{10} - 10^{11} \Msun$, which has been noted in several works \citep[e.g.,][]{Simpson:2018, Sales:2022, Riley:2025}.
\SatGen predicts fewer classical satellites ($\sim$10-15) than the other simulations, with a modest dependence on the adopted central disk mass.
We also note that the \SatGen results have larger scatter across all masses than any of the other simulation suites.
Given that our \SatGen results are based on 500 haloes, we expect this is a more accurate representation of the halo-to-halo scatter for a given model than the more limited samples (6 or 7 haloes) from the other suites.

For ultra-faints ($M_\star^\text{bound} < 10^5 \Msun$) in Aquarius and \SatGen, there is interesting agreement in the cumulative number of satellites within the $M_\star^\text{bound} \sim 10^3 - 10^4 \Msun$ range, but the two profiles exhibit different behavior at higher and lower masses.
The \SatGen results have a constant slope and the largest number of faint satellites, reflecting the adopted stellar mass-halo mass relation with no cutoff for forming galaxies in the lowest-mass haloes.
Comparing versions with different central stellar disks, the strongest reduction in the total number of satellites comes from introducing a disk at all, while doubling the mass of the disk does not have as much of an impact.

Meanwhile, Aquarius exhibits a bimodal distribution of faint and bright satellites, with peaks near $M_\star^\text{bound} \sim 3 \times 10^3 \Msun$ and $\sim 3 \times 10^6 \Msun$ and a dearth of satellites with $M_\star^\text{bound} \sim 10^5 \Msun$.
This is driven by the reionization physics implemented in GALFORM, specifically the parameters that control when reionization takes place and characteristic halo mass scale below which galaxy formation is suppressed \citep{Bose:2018}.
The suppression of galaxy formation in low-mass haloes results in a divergence between the Aquarius and \SatGen counts at the faintest end.
We note that the GALFORM model that underpins the Aquarius predictions is a modified version of that presented in \citet{Bower:2006} and newer versions of tagged Aquarius haloes with the latest GALFORM parameters produce similar results (Riley et al.~in prep).

\subsection{Bound fractions for classicals}

In Figure~\ref{fig:fbound-classicals} we present the distribution of $f_\text{bound}$ for classical-mass satellites across all of the simulation suites.
As we show the cumulative distribution, simulations closer to the lower right corner of the figure ($f_\text{bound} \sim 1$, $P \sim 0$) have the least disrupted satellite populations.
For visual clarity, we only show the scatter for the \SatGen run with $M_\text{d} / M_\text{vir} = 0.1$.
As supplementary material, we provide a figure that has the same information with simulations shown in separate panels, showing individual hosts (for Auriga, FIRE, and Aquarius) and the scatter for all \SatGen models.

The result is striking -- most models \textit{essentially agree, within their host-to-host scatter, across nearly the full range of disruption} (Auriga is the exception, which we discuss in a moment).
Between $\sim$20 and 30~per~cent of satellites have lost at least 10~per~cent of their stellar mass ($f_\text{bound} \leq 0.9$), while $\sim$10~per~cent of satellites have lost half or more of their stellar mass ($f_\text{bound} \leq 0.5$).
This is in spite of FIRE, Aquarius, and \SatGen sharing very few modeling choices among all three, beyond grounding their structure formation within a $\Lambda$CDM cosmology (though with slightly different adopted parameters).

On closer inspection, a number of interesting features emerge.
The \SatGen models with no stellar disk and $M_\text{d} / M_\text{vir} = 0.05$ are incredibly close to each other, while the $M_\text{d} / M_\text{vir} = 0.1$ model yields the expected result that increasing the stellar disk mass results in more stellar mass loss.
The $M_\text{d} / M_\text{vir} = 0.1$ model is also very close to the FIRE result.
We note that the present-day $M_\star / M_\text{vir}$ ratio ranges from 0.05 to 0.09 for FIRE and 0.06 to 0.07 for Auriga, with Auriga disks forming earlier in cosmic time \citep{Belokurov:2022}.
For the Milky Way, this value is likely between 0.02 and 0.05, with uncertainties in both the stellar \citep{Imig:2025, Lian:2025} and dark matter \citep[e.g.,][]{Wang:2020} components.
Aquarius acts as an interesting middle ground between these cases, agreeing with FIRE and the massive \SatGen disk results near $f_\text{bound} \sim 0.5$ and with the other \SatGen models for $f_\text{bound} \geq 0.9$.

The exception to the broad agreement across models is Auriga, which has a much more disrupted satellites.
For this suite, $\sim$60~per~cent of satellites have $f_\text{bound} \leq 0.9$, while nearly 40~per~cent of satellites have $f_\text{bound} \leq 0.5$.
While previous analysis of Auriga satellites led to the provocative claim that `most satellites are streams' \citep{Riley:2025, Shipp:2025}, it is clear that this result is not universal among models.
However, while Auriga disagrees with other simulations on \textit{the amount} of disruption that the Milky Way satellites should be experiencing, it agrees that the disruption is higher than conventionally expected or observationally detected (so far).
In Section~\ref{sec:conclusions} we discuss potential reasons for this enhanced disruption.

\subsection{Bound fractions for ultra-faints}

\begin{figure}
    \centering
    \includegraphics[width=\linewidth]{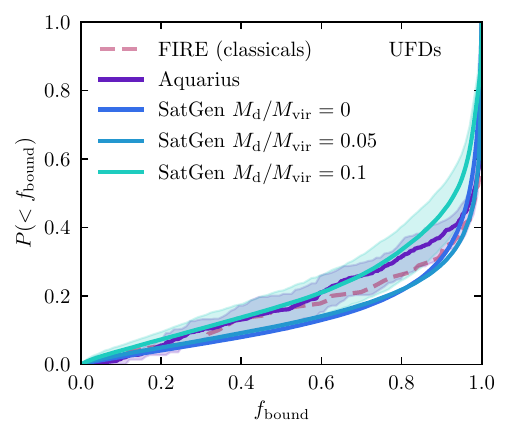}
    \caption{The cumulative distribution of $f_\text{bound}$ for ultra-faint satellites, shown in a similar manner as Figure~\ref{fig:fbound-classicals}.
    We only include systems that have a bound stellar mass of $10^2 \Msun < M_\star^\text{bound} < 10^5 \Msun$.
    The result for FIRE classical-mass satellites is also shown for comparison.
    \textit{Ultra-faint satellites experience stellar mass loss comparable to, or exceeding, those for classicals.}
    }
    \label{fig:fbound-ufds}
\end{figure}

In Figure~\ref{fig:fbound-ufds} we show the equivalent $f_\text{bound}$ distributions for ultra-faint satellites.
We only show results for Aquarius and \SatGen, which are the models with sufficient numerical resolution and physical prescriptions to analyze satellites with $M_\star^\text{bound} < 10^5 \Msun$.

The qualitative trends for ultra-faints are surprisingly similar to those for classicals.
The \SatGen models with no or lower disk mass are incredibly similar to each other, while the more massive disk model results in a more disrupted satellite population.
Aquarius also agrees with the \SatGen higher disk mass model for $f_\text{bound} \leq 0.7$ and transitions to agree with the other models for $f_\text{bound} \geq 0.9$.
The scatter between Aquarius and \SatGen are comparable, potentially indicating that improved statistics (many more ultra-faints than classicals) makes the halo-to-halo scatter for $f_\text{bound}$ distributions converge with fewer haloes.

Perhaps the most suprising result is that, for a fixed model, the ultra-faints show higher levels of disruption than the classicals.
Between $\sim$30 and 45~per~cent of satellites have $f_\text{bound} \leq 0.9$, while $\sim$10 to 20~per~cent of satellites have $f_\text{bound} \leq 0.5$.
We stress that though Aquarius and \SatGen have different galaxy-halo connections resulting in quite different predictions for the total number of ultra-faints (Figure~\ref{fig:luminosity-functions}), their $f_\text{bound}$ distributions are much more similar.

\section{Discussion and concluding remarks} \label{sec:conclusions}

It has been widely noted that satellites in cosmological (typically hydrodynamical) simulations are experiencing tidal disruption, even invoking this as a mechanism to alleviate small-scale problems of $\Lambda$CDM \citep{Brooks:2014, Wetzel:2016, Fattahi:2018}.
Several studies using controlled experiments have claimed that such results are subject to numerical effects \citep{vandenBosch:2018, Ludlow:2019, Ludlow:2023, Errani:2021, Chiang:2026} and that the results from hydrodynamical simulations in particular should be treated with caution.

Our baseline result -- that several models and methodologies all produce disrupting satellite populations -- is a substantial deviation from this conventional wisdom.
It is worth noting the numerical strengths and weaknesses of each simulation in turn.\footnote{We necessarily are painting with a broad brush, without commenting on whether these weaknesses are dealbreakers (or if the strengths are sufficient) in the context of disrupting satellites.}
Auriga and FIRE are likely most impacted by particle resolution and spurious heating, but include `messy' cosmological effects like a turbulent assembly history and satellite-satellite interactions.
Aquarius is not subject to spurious heating and is higher resolution than Auriga and FIRE, but does not include additional disruption due to a central stellar disk.
\SatGen does not suffer from numerical noise and can include the growth of a host disk (in a relatively simplified manner), but does not capture satellite-satellite interactions.
Despite this mixture of strengths and weaknesses, three of these models (FIRE, Aquarius, \SatGen) not only qualitatively agree that their satellites have experienced disruption, but on the \textit{degree of that disruption}.

The remaining model (Auriga) predicts a substantially higher level of disruption than the others.
As noted in Section~\ref{sec:results}, the stellar mass-halo mass relation for Auriga satellites is higher than comparable simulations and observational constraints (see Section~6.4 of \citealp{Riley:2025} for a summary).
Low-mass field galaxies in Auriga are also larger than observations at a fixed stellar mass, particularly for $M_\star < 10^8 \Msun$ \citep[e.g., Figure 1 of][]{Bose:2019}.
The Auriga central galaxies also mostly fall above the the typical $M_\star - M_\text{200c}$ relation \citep[e.g., Figure 10 of][]{Fragkoudi:2025}.
Taken together, this means that the typical $\sim$$10^{10} \Msun$ halo is populated with a more massive and more extended galaxy (larger satellites lose stellar mass more readily; Reddy et al.~in prep), then falls onto the main Auriga host to evolve in a stronger tidal field.
This combination could provide a physical explanation of the enhanced disruption relative to FIRE, even though the two share many of the same potential numerical effects.

It is worth considering that both Aquarius and \SatGen find that \textit{ultra-faints are more disrupted than classicals}.
Efforts to model Milky Way satellites typically assign stellar masses (or luminosities) to subhaloes based on their peak halo mass, then assume that this mass does not change unless the satellite is fully disrupted and removed from the sample \citep[e.g.,][]{Nadler:2020mwsats, Manwadkar:2022, Ahvazi:2025}.
This decision is typically justified by comparing notes with high-resolution numerical experiments.
However, we note that the tidal tracks used in \SatGen are calibrated to similar results \citep{Errani:2018}, suggesting that substantial stellar mass loss is entirely possible with such frameworks.
We encourage such efforts to not only consider increases in galaxy size from tidal heating (which are typically implemented), but also stellar mass loss in their forward modeling.

Stellar mass loss of ultra-faints has been relatively unexplored, since these systems are difficult to model in a Milky Way-like environment.
Taking the most conservative of these models (\SatGen with no disk) implies that of the 33 spectroscopically-confirmed ultra-faint galaxies discovered to date \citep{Pace:2025}, on average $\sim$10 have lost 10~per~cent of their stellar mass and $\sim$3 have lost over half of their stellar mass.
Meanwhile, only Tucana~III exhibits clearly detected tidal tails \citep{Li:2018}, while claims surrounding Bo\"{o}tes~III's tidal tails \citep{Carlin:2018, Jensen:2026, Li:2026} and nature as a galaxy or star cluster \citep{Zaremba:2025, Cerny:2026} are disputed.
If these models are correct, many ultra-faints have tidal tails yet undetected.
Recent detections \citep{Jensen:2024} of extended stellar distributions surrounding 7 Milky Way ultra-faint satellites (including Bo\"{o}tes~III and Tucana~III) may be an encouraging sign.
We note that diagnostics of disruption used in the literature (e.g., shape, size, proper motion alignment) may be unreliable indicators of a satellite's present-day disruption state \citep{Thoron:2026}.

We close on the following -- a variety of theoretical models predicting tidal disruption does not mean that these predictions reflect reality.
It is even possible that the recently unravelled debris could be at such a low surface brightness to never be detectable, even with upcoming facilities including NSF-DOE Rubin, \textit{Euclid}, \textit{Roman}, and \textit{Arrakihs}.
Detailed mock observations, in both the resolved star \citep{Shipp:2023} and integrated light \citep{Miro-Carretero:2025} regimes, will play a crucial role in determining if satellite disruption rates in the Milky Way and external galaxies are consistent with these models.
It is even possible that all of these models are wrong, requiring either new numerical prescriptions or even new physics that shield satellites from forming tidal streams and other debris.
However, the alternative is thrilling -- an abundance of low surface brightness tidal substructures that will reveal itself within the next decade, if galactic archeologists search for it.

\section*{Acknowledgements}
We thank Andrew Pace and Sarah Pearson for enlightening discussions.
AHR thanks Taylor Swift's \textit{The Life of a Showgirl} and Phoebe Bridgers's \textit{Stranger in the Alps} for providing a soundscape condusive to writing this manuscript.
This research made extensive use of the Science Explorer, funded by NASA under Cooperative Agreement 80NSSC21M00561, and \href{https://arxiv.org/}{arXiv.org} for bibliographic information.

AHR was supported by a fellowship funded by the Wenner Gren Foundation, a Research Fellowship from the Royal Commission for the Exhibition of 1851, and by STFC through grant ST/T000244/1.
AD is supported  by the European Union (ERC, BeyondSTREAMS, 101115754) grant.
Views and opinions expressed are however those of the authors only and do not necessarily reflect those of the European Union or the European Research Council.
Neither the European Union nor the granting authority can be held responsible for them.
OA acknowledges support from the Knut and Alice Wallenberg Foundation, the Swedish Research Council (grant 2025-04892), the Swedish National Space Agency (SNSA grants 2023-00164 and 2025-00405), the LMK foundation, and eSSENCE, a Swedish strategic research programme in e-Science.
SR-F acknowledges support from the SNSA (grant 2025-00181).
We acknowledge support from the DiRAC Institute in the Department of Astronomy at the University of Washington.
The DiRAC Institute is supported through generous gifts from the Charles and Lisa Simonyi Fund for Arts and Sciences, Janet and Lloyd Frink, and the Washington Research Foundation.

This work used the DiRAC@Durham facility managed by the Institute for Computational Cosmology on behalf of the STFC DiRAC HPC Facility (\href{https://dirac.ac.uk/}{www.dirac.ac.uk}).
The equipment was funded by BEIS capital funding via STFC capital grants ST/K00042X/1, ST/P002293/1, ST/R002371/1 and ST/S002502/1, Durham University and STFC operations grant ST/R000832/1.
DiRAC is part of the National e-Infrastructure.

For the purpose of open access, the author has applied a Creative Commons Attribution (CC BY) licence to any Author Accepted Manuscript version arising from this submission.

\section*{Data Availability}
We have used simulations from the Auriga Project public data release \citep{Grand:2024} available on the Globus platform.\footnote{\href{https://wwwmpa.mpa-garching.mpg.de/auriga/data.html}{https://wwwmpa.mpa-garching.mpg.de/auriga/data}}
We use the publicly-available FIRE-2 cosmological zoom-in simulations \citep{Wetzel:2023}, from the Feedback In Realistic Environments (FIRE) project, generated using the \textsc{Gizmo} code \citep{Hopkins:2015} and the FIRE-2 physics model \citep{Hopkins:2018}.\footnote{\href{https://flathub.flatironinstitute.org/fire}{https://flathub.flatironinstitute.org/fire}}
The Aquarius data is publicly available on Zenodo\footnote{\href{https://zenodo.org/records/13888986}{https://zenodo.org/records/13888986}} and detailed in \citet{Pu:2025}.
The \SatGen code is publicly available.\footnote{\href{https://github.com/shergreen/SatGen}{https://github.com/shergreen/SatGen}}

High-level catalogues produced in this work (stellar masses, bound fractions) will be made publicly available upon journal acceptance.

\section*{Software}
This research made use of the Python programming language, along with many community-developed or maintained software packages including:
    \textsc{astropy} \citep{Astropy:2013, Astropy:2018, Astropy:2022},
    \textsc{cmasher} \citep{cmasher, cmasher1.8.0},
    \textsc{cython} \citep{cython},
    \textsc{h5py} \citep{h5py, h5py3.7.0},
    \textsc{jupyter} \citep{ipython, jupyter},
    \textsc{matplotlib} \citep{matplotlib},
    \textsc{nbody\_streams} \citep{Arora:2026nbody_streams},
    \textsc{numpy} \citep{numpy},
    \textsc{pandas} \citep{pandas, pandas1.5.0},
    \textsc{scikit-learn} \citep{scikit-learn, scikit-learn-api, scikit-learn1.1.2},
    and \textsc{scipy} \citep{scipy, scipy1.9.1}.
We also thank the maintainers of the \textsc{arepo-snap-util} package.
Parts of the results in this work make use of the colormaps in the \textsc{cmasher} package.
Software citation information aggregated using \href{https://www.tomwagg.com/software-citation-station/}{The Software Citation Station} \citep{software-citation-station-paper, software-citation-station-zenodo}.

\bibliographystyle{mnras}
\bibliography{main, software}

\appendix

\section{Supplemental figure} \label{app:sim-panels}

\begin{figure*}
    \centering
    \includegraphics[width=\linewidth]{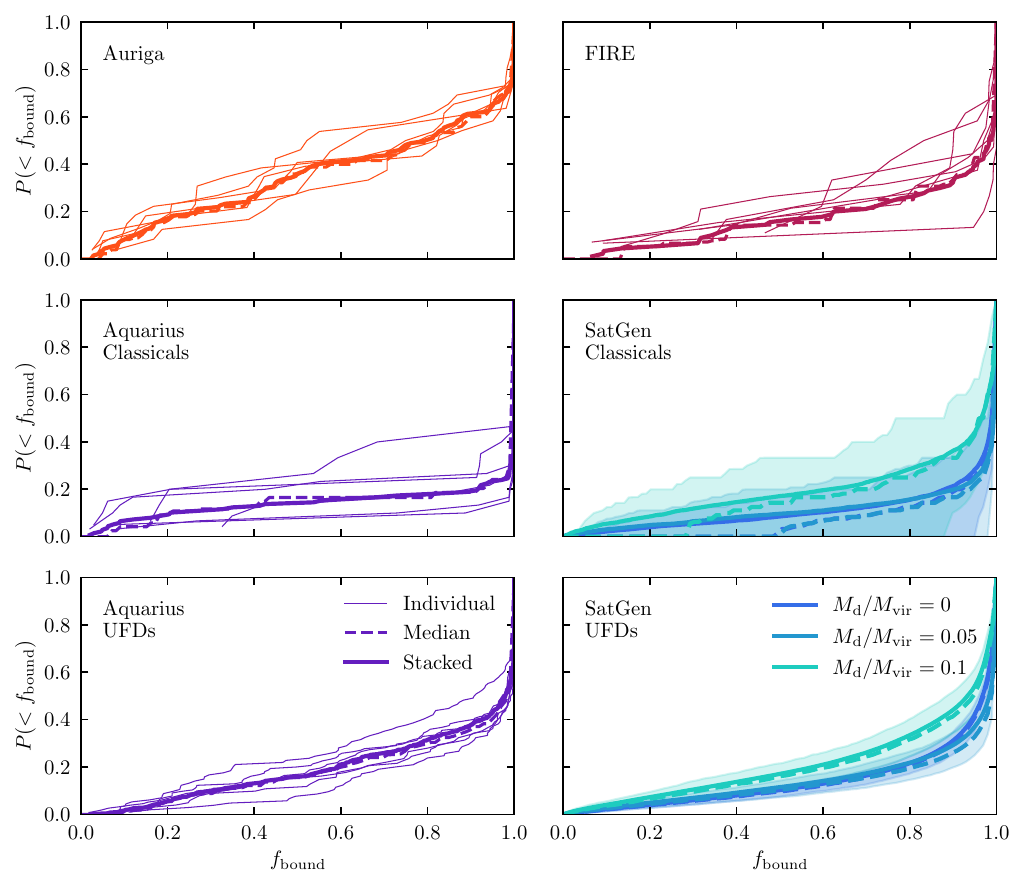}
    \caption{The cumulative distribution of $f_\text{bound}$ for satellites across different simulation setups.
    The information is the same as is presented in Figures~2 and 3, but each panel corresponds to a different simulation.
    Simulations that produce particle distributions (Auriga, FIRE, Aquarius) have each individual haloes shown as a thin lines.
    The \SatGen realizations are shown as shaded contours, corresponding to the 1$\sigma$ percentiles across all 500~haloes.
    The medians across all haloes are shown as thick dashed lines, while the stacked distributions (what is shown in the main text) are shown as thick solid lines.
    \textit{Despite substantial differences the galaxy-halo connections and other modeling choices, all of the simulations predict substantial stellar mass loss for satellites of Milky Way-mass haloes.}
    }
    \label{fig:fbound-sim-panels}
\end{figure*}

In Figure~\ref{fig:fbound-sim-panels} we show the $f_\text{bound}$ distributions with simulations separated by panel to aid visual comparison.
This contains the same data as presented in Figures~\ref{fig:fbound-classicals} and \ref{fig:fbound-ufds}.

\bsp	
\label{lastpage}
\end{document}